\documentclass[aps,preprint]{revtex4}

\begin{document}

\title{Multi-quark states}
\author{Fan Wang}
\affiliation{Department of Physics and Center for Theoretical Physics,
Nanjing University, Nanjing, 210093, China\\}
\author{Jia Lun Ping}
\affiliation{Department of Physics, Nanjing Normal university, Nanjing, 210097
and Center for Theoretical Physics, Nanjing University, Nanjing, 210093, China}
\author{Di Qing}
\affiliation{CERN, Ch-1211 Geneva 23, Switzerland}
\author{T. Goldman}
\affiliation{Theoretical Division, LANL, Los Alamos, NM87545, USA}
\date{May 2004, a tribute to the seventy birthday of Yu. A. Simonov}

\begin{abstract}
The pentaquark state recently discovered has been discussed based
on various quark model calculations. Odd parity for the state can
not be ruled out theoretically because contributions related to
non-trivial color structures have not been studied completely.
Other multiquark states, especially dibaryons, have been discussed
also. A strangeness -3 N$\Omega$ dibaryon has been shown to have a
width as small as 12-22 keV and should be detectable in $\Omega$
high productivity reactions such as at RHIC, COMPAS and the
planned JHF and FAIR projects.
\end{abstract}
\maketitle

\bigskip

\section{multiquark state search}
Multi-quark states are studied even before the advent of QCD. The
development of QCD accelerated multi-quark studies because it is
natural in QCD that there should be multi-quark states, including
glueballs and quark-gluon hybrids. Prof. Yu. A. Simonov is one of
the pioneers of multi-quark studies; he led an ITEP group that
developed the quark compound bag model in the early 1980's to
study hadron interactions and multiquark states\cite{yu}.

For a long time, multi-quark states were only a theoretical speculation;
experimental searches had not obtained definite evidence, even though
there were various claims of the discovery of multi-quark states.

\section{discovery of the pentaquark}
Eleven groups\cite{exp} claimed recently that they found a
pentaquark state, now called $\Theta^+$, with mass$\sim$1540 MeV,
and width $\Gamma < 25$ MeV. Five measurements used a real or
virtual photon-nucleus (p, d, or other nuclei) reaction and the
resonance is inferred from the final state $nK^+$ or $pK^0_s$
invariant mass. A reanalysis of 1986 bubble chamber K-nucleus
reaction data also found the $\Theta^+$. Three most recent
experiments used pp or p-nucleus reactions. One used old neutrino
reaction data and one used deep inelastic {\em ep} scattering
data. In addition, the NA49 collaboration claimed that they found
the anti-decuplet partner $\Xi^{--}$ of $\Theta^+$\cite{na49}. The
HERA-H1 collaboration claimed that they found the charm pentaquark
$\Theta_c$\cite{h1}.

If we had not had bitter previous experience in the search for
multiquark states, one would have accepted that the pentaquark
state has been discovered. Taking into account the historical
lessons that many low statistics multiquark signals eventually
disappeared, the consensus is that high statistics data are
necessary to confirm this state.

Moreover there are weaknesses among these measurements: Some
"pentaquark signals" of the real or virtual photon reactions
might be due to the kinematical reflection of the following
normal meson production processes\cite{dzi}
\begin{eqnarray}
\gamma n\rightarrow f_2n\rightarrow K^-K^+n,  \nonumber\\
\gamma p\rightarrow a_2^+n\rightarrow K\bar{K}n, \nonumber\\
\gamma p\rightarrow f_2p\rightarrow K\bar{K}p.
\end{eqnarray}
Some measurements only measure the charged $\pi^+,\pi^-$ decay
products of $K^0_s$ to obtain the p$K^0_s$ invariant mass and
there might be kinematical reflections in these events
too\cite{zav}. Reanalysis of the vast $K^+n$ and $K^0_Lp$
scattering data could not find the $\Theta^+$ signal with a width
$\ge$ few MeV\cite{arn}, except the recent time delay and speed
plot analysis where a signal appears around 1.57 GeV in the
$P_{01}$ (with the notation $L_{I(2J)}$) channel\cite{kel}. In
addition, a very tiny bump had appeared in 1973 CERN
$K^+p\rightarrow pK^0_s\pi^+$ inelastic scattering data\cite{les}.

The NA49 claim has been challenged by another CERN group based on $\Xi$
spectroscopy data with higher statistics\cite{fis}.  HERA-B p-nucleus
reaction data has not found the $\Theta^+$\cite{herab}. BES $J/\Psi$
decay data analysis has not found the $\Theta^+$ either\cite{bes}.
There are other groups that have not claimed publicly that they have
not found the $\Theta^+$ signal.

\section{what is the pentaquark}
Theoretical studies based on the chiral soliton model played an
important role in triggering the $\Theta^+$ searches\cite{dia}. In
the chiral soliton quark model, the $\Theta^+$ is a member of the
anti-decuplet rotational excitation following the well established
octet and decuplet baryons. It predicted an I=0
$J^p=\frac{1}{2}^+$ state with mass about 1540 MeV and width less
than 15 MeV, quite close to the later experimental results. The
QCD background of this model has been criticized by Jaffe and
Wilczek\cite{jw} and the difficulties of the flavor SU(3)
extension in the description of the meson-baryon scattering with
strangeness was discussed by Karliner and Mattis\cite{km}.

Various quark models have been proposed to understand the $\Theta^+$,
mainly aimed at explaining the low mass and narrow width which is not
a serious problem in the chiral soliton quark model but hard to understand
in the usual quark model. First the ground state should be an S-wave
one in the naive quark model\cite{rgg,isg} and this means $\Theta^+$
should be a negative parity state. The S-wave $uudd\bar{s}$
configuration will have S-wave KN components. However both the I=0,1
KN S-wave phase shifts are negative in the $\Theta^+$ energy region
and this means that $\Theta^+$ can not be an S-wave KN resonance.
On the other hand, since the I=0 KN P-wave $P_{01}$ phase shifts are
positive, there might be resonance in this channel and this is
consistent with the $J^p=\frac{1}{2}^+$ predicted by the chiral
soliton quark model of the spin-parity of $\Theta^+$. If the
$\Theta^+$ is confirmed to be a positive parity state and there is
no negative parity pentaquark state with lower energy then it will
provide another example of an inverted energy level structure which
has been a weak point of the naive quark model of the baryon spectrum:
The first excited states of N and $\Delta$ have positive parity
instead of negative parity as predicted by the naive quark model.

Jaffe and Wilczek\cite{jw} proposed a $\{ud\}\{ud\}\bar{s}$
structure for the pentaquark state, where $\{ud\}$ means a
di-quark in the color anti-triplet, spin 0, isospin 0 S-wave
state. They assume a strong color-spin force to argue for such a
di-quark structure, a color Cooper pair. The overall color singlet
condition dictates that the four quarks must be in a color triplet
state (in a color SU(3) [211] representation), which is
antisymmetric with respect to the di-quark exchange in color
space. The two quasi-boson di-quarks should be symmetric with
respect to the overall di-quark exchange and this dictated the
relative motion of the two di-quarks in a P-wave. In this way they
get a positive parity $\Theta^+$ state, the same as the chiral
soliton quark model one. This model gives an antidecuplet spectrum
different from the chiral soliton quark model ones and predicts a
pentaquark octet based on phenomenological assumptions. Yu. A.
Simonov did a quantitative calculation based on the Jaffe-Wilczek
configuration by means of the effective Hamiltonian approach. This
method obtained a quantitative fit of single baryon masses, but
the calculated $\Theta^+$ mass is about 400 MeV higher than the
observed 1540 MeV\cite{yua}. F. Close pointed out that there
should be a spin-orbit partner $\Theta^*$ with $J^P=\frac{3}{2}^+$
not too far from the $\Theta^+$ mass\cite{clo}.

An even more serious problem is the decay width. F. Buccella and
P. Sorba suggested an approximate flavor symmetry selection rule
to explain the narrow width of $\Theta^+$ based on the
Jaffe-Wilczek model\cite{bs}. However if the Jaffe-Wilczek
correlated di-quark di-quark wave function is totally
anti-symmetrized then there will still be an SU(6) totally
symmetric component and no flavor symmetry selection rule to
forbid the pentaquark to decay to a KN final state. Diakonov and
Petrov even raised the criticism that the non-relativistic quark
model might not make any sense for light flavor pentaquark
states\cite{dip}.

Fl. Stancu, D.O. Riska and L.Ya. Glozman gave another explanation
of the special properties of the pentaquark state using their
Goldstone boson exchange quark model\cite{srg}. The flavor-spin
dependence of the Goldstone boson exchange qq interaction favors a
totally symmetric flavor-spin wave function of the four light
quarks. The overall color singlet dictates the four light quarks
to be in a color triplet state, i.e., in a color SU(3) [211]
representation which is the same as discussed in the Jaffe-Wilczek
model. The Pauli principle further dictates the four light quarks
to be in a [31] representation of SO(3), therefore they must be in
a $s^3p$ configuration. The kinetic energy increase might be
compensated by a potential energy decrease. However the Goldstone
boson exchange interaction itself might not be enough to form the
$\Theta^+$ and they have to introduce an additional $q-\bar{s}$
interaction due to $\eta$ meson exchange which was argued not to
exist in this model approach\cite{ns}. The narrow width of
$\Theta^+$ has not been discussed.

B.K. Jennings and K. Maltman did a comparative study of the three
models of the $\Theta^+$ mentioned above and related exotics\cite{jm}.
They concluded that the three models appear to be different but might
describe the same physics. They also pointed out that the narrow
width of $\Theta^+$ may be explained by the small overlap of the five
quark model wave function and the KN one. But they avoid discussion
of the mass of the pentaquark because that is dependent on estimates
of the confinement and kinetic energy.

Our group has done three quark model calculations. The first one is an
application of the Fock space expansion model which we developed to
explain the nucleon spin structure\cite{qcw}. The naive quark model
assumes that the baryon has a pure valence $q^3$ configuration. This
is certainly an approximation. One expects there should be higher
Fock components,
\begin{equation}
B = aq^3 + bq^3q\bar{q}+\cdots.
\end{equation}
The nucleon spin structure discovered in polarized lepton-nucleon
deep inelastic scattering can be explained by allowing the nucleon
ground state to have about 15$\%$ $q^3q\bar{q}$ component where the
$q\bar{q}$ parts have pseudoscalar meson quantum numbers. In the
$\Theta^+$ mass calculation we assume it is a pure $uudd\bar{s}$
five quark state but with channel coupling. Our preliminary results
are listed below:
\begin{eqnarray}
&&~~pure KN~~~~KN+K^*N~~~~KN+K_8N_8~~~~KN+K^*N+K_8N_8 \nonumber\\
&&I. S_{01}~~ negative~~parity \nonumber \\
&&~~~~2282~~~~~~~~~~2157~~~~~~~~~~~~~1943~~~~~~~~~~~~~~~~1766        \nonumber\\
&&II. P_{01}~~positive~~parity \nonumber \\
&&~~~~2357.1~~~~~~~~2356.3~~~~~~~~~~2357.0~~~~~~~~~~~~~~~2336.8 \nonumber
\end{eqnarray}
Here the calculated mass is in units of MeV and $K_8N_8$ means the
K and N are both in color octets but coupled to an overall color
singlet. These numerical results are under further check. However
the following results will not change: The S-wave state will
have a lower mass than that of the P-wave; the channel coupling
plays a vital role in reducing the calculated S-wave $\Theta^+$ mass;
it is possible to obtain a mass close to the observed mass 1540 MeV
of the $\Theta^+$ if the overestimation of the K mass in this model
is taken into account.

Quite possibly, confinement is due to gluon flux tube (or gluon
string) formation in the QCD vacuum. Yu. A. Simonov proposed a
field correlator method to study the non-Abeliean non-perturbative
properties of QCD and found a Y-shaped gluon flux tube (or string),
but not a $\Delta$-shaped one, in a baryon. The ground state
energy can be approximated by a potential\cite{yu88}
\begin{eqnarray}
&&V_{3q}=-A_{3q}\sum_{i<j}\frac{1}{|\vec{r}_i-\vec{r}_j|}
        +\sigma_{3q}L_{min}+C_{3q},  \nonumber\\
&&\L_{min}=\sum_i L_i.
\label{pot}
\end{eqnarray}
$L_i$ is the distance between the quark i and the Y-shaped gluon
junction. $\vec{r}_i$ is the position of quark i. The first term
in Eq.(\ref{pot}) is the color Coulomb interaction and the second
term is similar to a linear confinement potential.

Most of constituent quark models use a quadratic or linear
potential to model the quark confinement,
\begin{eqnarray}
V_{conf}(\vec{r}_{ij})=-a\vec{\lambda}_i\cdot\vec{\lambda}_j\vec{r}^n_{ij},
\nonumber\\
\vec{r}_{ij}=\vec{r}_i-\vec{r}_j, ~~~~~~   n=1,2.
\label{conf}
\end{eqnarray}
Here $\lambda^a_i$ ($a=1\cdots 8$) is the color SU(3) group
generator. For a single hadron, $q\bar{q}$ mesons or $q^3$
baryons, such a modelling can be achieved by adjusting the
strength constant $a$ of the confinement potential. The color
factor $\vec{\lambda}_i\cdot\vec{\lambda}_j$ gives rise to a
strength ratio $1/2$ for baryon and meson which is almost the
ratio for the minimum length of the flux tube to the circumference
of the triangle formed by three valence quarks of a baryon.

How to extend the confinement potential to multiquark systems is
an open question. There are a few lattice QCD calculations of
pentaquarks which obtained an S-wave ground state\cite{lat}.
However they have not given the color flux tube or string
structure as the Simonov group did for mesons, baryons and
glueballs.  But from general SU(3) color group considerations,
there might be the following color structures:
${q^3}(1){q\bar{s}}(1)$; ${q^3}(8){q\bar{s}}(8)$;
${qq}(\bar{3}){qq}(\bar{3})\bar{s}(\bar{3})$, etc. Here the number
in parentheses is the color SU(3) representation labelled by its
dimensions. The first one is a color singlet meson-baryon channel;
the second is the hidden color meson-baryon channel; the third is
the color structure used in the Jaffe-Wilczek model. New color
structures will give rise to additional interactions which have
not been taken into account in the quark model calculations of the
pentaquark so far. And our first model calculation mentioned above
shows that hidden color channel coupling reduces the calculated
pentaquark mass.

We have developed a model, called the quark delocalization, color
screening model (QDCSM), to take into account the additional
interaction in multiquark systems induced by various color
structures, which are not possible for a $q\bar{q}$ meson and
$q^3$ baryon, by a re-parametrization of color
confinement\cite{wang1}. This model explains the existing BB
interaction data (bound state deuteron and NN, N$\Lambda$,
N$\Sigma$ scatterings) well with all model parameters fixed in
hadron spectroscopy except for only one additional parameter, the
color screening constant $\mu$. More importantly, {\it it is the
unique model, so far, which explains a long-standing fact: The
nuclear force and the molecular force are similar except for the
obvious difference of length and energy scales; the nucleus is
approximately an A nucleon system rather than a 3A quark
system}\cite{wang2}.

A preliminary calculation of the pentaquark mass has been done
with this model (QDCSM) in the color singlet KN configuration. In
the I=1 S-wave KN channel, a pure repulsive effective interaction
is obtained. This helps to rule out the I=1 possibility for the
$\Theta^+$. In the I=0 S-wave KN channel, a $\Theta^+$ mass of
1615 MeV is obtained in an adiabatic approximation. More precise
dynamical calculation might reduce the calculated mass further.
For the P-wave channels, we only obtain spin averaged effective KN
interactions because the spin-orbit coupling has not been included
yet. In the I=0 channel, there is a strong attraction, as wanted
in other quark models. However in our model the P-wave attraction
is not strong enough to overcome the kinetic energy increase to
reduce it to a ground state. This is consistent with the lattice
and QCD sum rule results\cite{lat,zhu}. In the I=1 channel, only a
very weak attraction is obtained.  This rules out the I=1
$\Theta^+$ again.

In a third model, we use the $\{ud\}\{ud\}\bar{s}$ configuration.
The color structure is the same as the Jaffe-Wilczek one but the
four non-strange quarks are totally anti-symmetrized. The space
part is fixed to be an equilateral triangle with the two diquarks
sitting at the bottom corners and the $\bar{s}$ at top. The height
and the length of the bottom side of the triangle are taken as
variational parameters in addition to the quark delocalization. A
three body variational calculation with the QDCSM has been done.
The minimum of this variational calculation is around 1.3-1.4 GeV
corresponding to a color screening parameter $\mu=1.0-0.8
fm^{-2}$. This gives rise to an effective attraction with a
minimum at around 50-150 MeV, qualitatively similar to our second
model result but with the possibility of further reducing the
calculated $\Theta^+$ mass to be closer to the observed value.

As mentioned before these numerical results have to be checked
further and the QDCSM is only a model of QCD. Based on these
results we cannot definitively claim that the ground state of
the $\Theta^+$ is $IJ^P=0\frac{1}{2}^-$. However, this possibility
has not been ruled out because color confinement contributions of
various nontrivial color structures for a pentaquark system have
not been studied thoroughly. This can be checked by devising
pentaquark interpolating field operators with these nontrivial
color structures in lattice QCD calculations.

Suppose the $\Theta^+$ is finally verified to be a 1540 MeV narrow
width ($\sim$ 1 MeV) $IJ^P=0\frac{1}{2}^+$ state. Then an
interesting scenario similar to that of nuclear structure at the
1940's turned to the 1950's will recur. The low lying, even parity
rotational excitation of nuclei is hard to explain by the naive
Mayer-Jenson nuclear shell model; Bohr and Mottelson had to
introduce the rotational excitation of a deformed liquid drop
model. Later, nucleon Cooper pairs were introduced because of the
strong short range pairing correlation. In 1970's-1980's, an S-D
Cooper pair interaction boson model was developed and the
collective rotation was re-derived from this model which is based
on Mayer-Jenson's nuclear shell model but with nucleon pair
correlation. In the description of the pentaquark, one has
introduced the chiral soliton rotational excitation, quark color
Cooper pairs and much more. The historical lessons of nuclear
structure study might be a good mirror to light the way for the
study of hadron structure.

\section{Tetraquarks, hybrids and glueballs}
By the end of 1970, Jaffe had already suggested that the scalar
mesons with masses less than 1 GeV ($\sigma$(600), $f_0$(980),
$a_0$(980), $\kappa$(900)) might be $qq\bar{q}\bar{q}$
states\cite{jaffe}. The "discovery" of a pentaquark enhances
this possibility. In addition there are a few new candidates for
tetraquarks: the BABAR experiment observed a resonance of
M=2317~MeV, $\Gamma$ $<$ 10 MeV in the $D_s^+ \pi^0$ invariant
mass analysis\cite{babar}. CLEO confirmed this resonance and
observed a new $D_{sJ}^+$(2463 MeV) state\cite{cleo}. These might
be $q\bar{q}c\bar{s}$ tetraquark states. Belle observed a resonance
of 3872 Mev in the $J/\Psi\pi^+\pi^-$ channel\cite{belle} in B meson
decay. CDF-II confirmed this resonance in $p\bar{p}$ collisions\cite{cdf}.
This might be a $D^0\bar{D^{*0}}$ molecular state.

There have been glueball and hybrid candidates. In general these states
will be mixed and also mixed with tetraquark states. The "discovery"
of a pentaquark state will make the identification of glueballs and
hybrids even harder\cite{kle}.

\section{Dibaryons}
Immediately after the development of MIT bag model, Jaffe predicted
the deeply bound dibaryon H, an I=S=0 uuddss flavor singlet state\cite{j77}.
Long term extensive searches for the H have been null. Yu.A. Simonov
proposed the quark compound bag model and various dibaryon resonances
were predicted\cite{nucl}. E.L. Lomon extended the R-matrix formalism
to dibaryon studies\cite{lomon}. This model explains the NN scattering
data well because it has the aid of the boundary condition model of the
NN interaction\cite{bound}. It predicted a 2.7 GeV NN resonance and an
experimental signal had been claimed.

A direct extension of the naive quark model\cite{rgg,isg} to
baryon-baryon (BB) interactions only gives rise to the short range
repulsion of the NN interaction but not the intermediate range
attraction nor the well established long range one $\pi$ exchange
tail in the meson exchange model. This means that important physics
has been missed in the application of the naive quark model to the
BB interaction. One possibility arises from spontaneously broken
chiral symmetry and the related Goldstone boson exchange\cite{mg}.
I.T. Obukhovsky and A.M. Kusainov applied such a hybrid model,
with both gluon and Goldstone boson exchange, to the NN
interaction\cite{obu}. Z.Y. Zhang et al. extended this hybrid model
from SU(2) to SU(3) and used it to predict dibaryon
resonances\cite{zhang}. Such a direct extension with a universal
$\sigma$-u,d,s quark coupling will overestimate the $\sigma$
meson contribution in high strangeness channels and makes their
predicted mass of the di-$\Omega$ too low\cite{ppw}.

V.B. Kopeliovich et al. studied the dibaryon with the Skyrmion model
and predicted dibaryons up to high strangness -6. But after taking into
account the Casimir energy, these dibaryons become unbound\cite{kop}.

We have studied the dibaryon candidates in the u, d, s three
flavor world systematically with both non-relativistic and
relativistic versions of the QDCSM in an adiabatic
approximation\cite{wg}. This model has been mentioned before. The
main new ingredients are: {\it Baryon distortion or internal
excitation in the course of interaction is included by allowing
the quark mutual delocalization between two interacting baryons
and a variational method is developed to let the interacting
baryon choose its own favorable configuration; a new
parametrization of color confinement is assumed to take into
account the contribution of nontrivial color structures which are
not possible in a q$\bar{q}$ meson and $q^3$ baryon.} Only a few
states remain after filtration with a more precise dynamic
calculation: A nonstrange $IJ^p=03^+$ $d^*$ (M=2165-2186 MeV,
$\Gamma$=5.76-7.92 MeV)\cite{ping} and a strangeness -3
$IJ^P=\frac{1}{2}2^+$ N$\Omega$ (M=2549 MeV, $\Gamma$=12-22 KeV).
The H and di-$\Omega$ are near threshold states and might be
unbound when the model uncertainty is taken into
account\cite{pang}. There are few broad N$\Delta$ and
$\Delta\Delta$ resonances with widths of 150 MeV-250 MeV in the
energy range 2.1-2.2~GeV which might make the identification of
the $d^*$ even harder and all of them might be the origin of the
observed broad resonance of the pp and np total cross section in
that energy region. The strangeness -3 N$\Omega$ state is a very
narrow dibaryon resonance and might be detected in high $\Omega$
productivity reactions such as at RHIC, COMPAS and the planned JHF
in Japan and FAIR in German.

\section{Summary}
Multi-quark states have been studied for about 30 years. The $\Theta^+$,
if further confirmed, will be the first example. Once the multi-quark
"Pandora's box" is opened, the other multi-quark states: tetra-quark,
hexa-quark (or dibaryon), etc., can not be kept inside. One expects
they will be discovered sooner or later. A new landscape of hadron
physics will appear and it will not only show new forms of hadronic
matter but will also exhibit new features of low energy QCD.

Nonpertubative and lattice QCD have revealed the color flux tube (or
string) structure of the $q\bar{q}$ and $q^3$ states. The multi-quark
system will have more color structures. How do these color structures
interplay within a multiquark state? Nuclear structure seems to be
understood in terms of colorless nucleons within a nucleus. We emphasized
that other color structures should be studied and the multi-quark systems
provide a good field to do that. The low mass and narrow width of the
$\Theta^+$ might be related to such new structures instead of to
residual interactions.

I apologize to those authors whose contributions in this field have
not been mentioned due to the limitation of space. This work is supported
by the NSFC under grant 90103018, 10375030 and in part by the US Department
of Energy under contract W-7405-ENG-36.

\end{document}